# PMU-Driven Non-Preemptive Disconnection of Overhead Lines at the Approach or Break-Out of Forest Fires

Panayiotis Moutis, *Senior Member, IEEE* and Uday Sriram

*Abstract–* The number and intensity of forest fires in the United States have been steadily increasing and causing massive economic and ecological damage. Overhead transmission and distribution lines are typically disconnected preemptively near fires, a practice that causes high social cost. Deployment of Phasor Measurement Units (PMUs) and their ability for synchronized real-time measurement of voltage and current inspire the here proposed method for the just-in-time disconnection of overhead lines at the ignition or approach of fires. PMUs can detect ambient temperature changes caused by fires through the change of line resistance. An ambient temperature model at the vicinity of a fire seat and the dynamic heat exchange model of the IEEE 738 dynamic line rating standard for bare overhead conductors are used in MATLAB Simulink to assess the performance of the proposed control method. Extensive testing shows that, under various conditions, the proposed technique can detect a forest fire approaching an overhead line, while avoiding, almost completely, any false alarms.

*Index Terms*—Fires, power distribution lines, PMU, protection, synchrophasor, transmission lines.

## NOMENCLATURE

| | |
|---|---|
| $A$ | Thermal absorption coefficient per meter |
| $cos\varphi$ | Power factor |
| $d$ | Distance |
| $I$ | Current |
| $I_e$ | Current measurement error |
| $l$ | Length of an overhead line |
| $m \cdot C_p$ | Total heat capacity |
| $n$ | Nominal value |
| $P_R, Q_R$ | Active and reactive power at receiving terminal $R$ |
| $q_c$ | Convection heat loss |
| $q_f$ | Thermal radiation from a fire |
| $q_r$ | Radiated heat loss |
| $q_s$ | Heat gain from solar irradiation |
| $R$ | Resistance |
| $T$ | Period of electrical cycle |
| $T_a$ | Ambient temperature |
| $T_c$ | Average temperature of the strand layers of a conductor |
| $T_s$ | Conductor surface temperature |
| $t$ | Time |
| $V_e$ | Voltage measurement error |
| $V_S, V_R$ | Voltage magnitude at sending $S$ and receiving $R$ terminals |
| $V_w$ | Wind speed |
| $X$ | Reactance |
| $Z$ | Impedance, where $Z=R+jX$ |
| $\alpha$ | Temperature coefficient of material |
| $\Delta$ | Change or rate of change |
| $\delta$ | Impedance angle in polar coordinates |
| $\theta$ | Voltage angle (phase) difference |
| $\lambda$ | Thermal conductivity |
| $\rho$ | Air density |

## I. INTRODUCTION

RECENTLY, forest fires have caused extensive environmental and economic damages in the United States (US). California (CA), specifically, and the broader Western US have suffered about 3500 fires, burning 7 million acres and causing damages of around $4.8 billion in the last 10 years [1]. Although dry weather conditions, strong winds (indications of climate change), and availability of fuel have been the main contributors to forest fires [2], around 84% of the forest fires between 1997-2017 were caused by humans, including electrical grid events [3].

Typically, system operators will preemptively disconnect power lines well in advance and in wide regions in the vicinity of fires. Although this practice protects wires from arcing faults (due to ash/smoke), damage or further ignition of additional fire seats [4, 5], it bears high social cost. During the 2019 CA fires, the local utility left over 940,000 houses and businesses without electricity (2.7 million people) [6]. Nevertheless, there is limited research that accounts for the effect of power systems to forest fires and vice versa. Authors in [7] analyze the effects of fires in four typical types of vegetation in the event of AC transmission line breakdown and typical prevention techniques. The fault-clearing reclosing in [8] takes into account when a line is in a fire hazard zone. An optimal power outage scheduling at the event of a wildfire is discussed in [9]. Some more focused research may be found in the detection of down conductors as they may be responsible for fire ignition [10]. Many studies attempt to analyze fire spread characteristics and improve fire safety methods, such as the work in [11], with geospatial interpolation techniques to map fire progression within a 5000 acre area in Northern CA.

This work proposes a control method for the timely (non-preemptive) disconnection of overhead lines at the approach of a fire front. The control is enabled thanks to Phasor Measurement Units (PMUs) [12] that monitor changes in line resistance

P. Moutis is with the Scott Institute for Energy Innovation, Carnegie Mellon University, Pittsburgh, PA 15213 USA (e-mail: pmoutis@andrew.cmu.edu).

U. Sriram was with the Dept. of ECE, Carnegie Mellon University, Pittsburgh, PA 15213 USA (e-mail: udaysriram97@gmail.com).

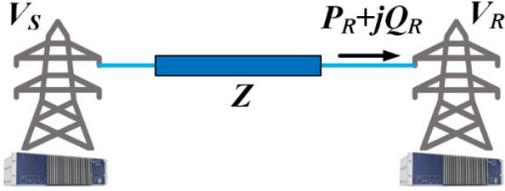

Fig. 1. PMUs at sending and receiving ends of an overhead line for control. For the proposed methodology, either of the PMUs may be controlling a breaker for the timely disconnection of the line at the approach of a fire.

caused by changes in ambient temperature, as a result of a fire nearby. A few studies propose similar approaches about the identification of line characteristics with PMU measurements, albeit not for the case of forest fires. In [13] a method inspired by state estimation identifies the parameters of transmission lines and transformers. A few measurement sets (adding to the time requirements of the iterative method) are needed to reduce the estimate error of the asset characteristics. The line parameter estimation in [14] uses a two-stage and a two-port line model to address asymmetries introduced by untrasposed lines and unbalanced load. At the presence of measurement error, this method estimates more accurately the characteristics of long lines. The work in [15] also focuses on the estimation of parameters of long lines, while it also attempts to calculate line temperature (and sag) based on a linear relationship of resistance to ambient temperature. A similar linear relationship is assumed also in [16]. Multiple measurement sets, albeit for a limited deployment of PMUs are used in [17], to identify line parameters in cases that measurements are scarce. PMU scarcity is handled in [18] with the joint use of SCADA measurements, while the focus is on long, multi-terminal lines. The authors of [17] improve their methodology in [19] by first identifying erroneous line parameter estimates and then improving the estimate with a day's data. In both these works, the authors assume that line parameters are constant. The work in [20] takes into account the errors introduced by phase drift to calculate the Thevenin impedance (equivalent of whole grid impedance) at the terminals of a generator. To handle measurement error, ambient temperature effects and asymmetries, [21] proposes a M-estimator. In [22] PMU measurements and detailed heat exchange models are used to estimate resistance and temperature of underground cables. The method requires hour-long data, and outputs estimates that may lie in ±10ºC (or more) confidence intervals every 5'.

The here presented work is also relevant to dynamic line rating (DLR) studies. DLR is a model seeking to assess in real-time (preferably) the loading limits of transmission lines, which differ from their static ratings [23, 24]. As DLR is pertinent to ambient conditions (solar irradiation, wind speed and temperature), heat exchange models inform these models. Various types of monitoring devices may also be installed on transmission lines [25] to further improve DLR models with direct measurements of their temperature and/or sag.

The here proposed method differs from the cited ones both in formulation and scope. To capture line resistance changes induced by increases of ambient temperature, due to a nearby fire, the ratio of the line reactance over resistance is used. That ratio is the tangent of the angle of the line impedance phasor, which we calculate with the synchronized PMU measurements of voltage and current at the ends of the line as shown in Fig. 1. In this sense, an estimator, either linear or iterative, is avoided. The *scope* of the method is in disconnecting lines near new fire seats, as those of newly breaking out fires or ignited by blasting pinecones far from a known fire-front [26]. These lines cannot be assumed to be long; a new fire seat will span short lengths and, thus, affect small parts of overhead conductors. Further, the control must act fast as system operators aim to deenergize assets to protect them from arcing faults induced by smoke and ash [4]. The *innovative contributions* of this work include:
- the detection of a fire seat near an overhead line in times of a few electrical cycles,
- a formulation for identifying some line characteristics with limited processing requirements (no use of estimators), thus, also allowing local control action,
- the realistic consideration of the effects of ambient temperature changes to line resistance through heat exchange models,
- an added value proposition for PMUs (which uniquely enable this method) and the avoidance of use of additional monitoring devices such as those purposed for DLR modeling,
- the prevention of the socially costly preemptive interruption of supply to end-customers, and
- the protection of assets from damages and faults by fire fronts or seats that that have not been anticipated.

Section II describes the proposed control method, an ambient temperature model at the vicinity of a fire seat and the heat exchange model for an overhead line. Extensive testing of the method in MATLAB Simulink over multiple operating scenarios is presented in Section III, which also includes an analysis of the efficiency of the control with machine learning tools. In Section IV the results are discussed and concerns for practical implementation are put forth. Section V concludes this work.

## II. PMU-DRIVEN DISCONNECTION OF OVERHEAD LINES THROUGH MONITORING OF AMBIENT TEMPERATURE CHANGES WITH THE LINE X/R RATIO

### A. Relationship of Line Resistance to Ambient Temperature

Resistance is directly proportional to temperature changes, increasing above, typically, 20ºC and vice versa for temperature below that [27], as per the following formula:

$$R(T_c) = R_{ref} \cdot [1 + \alpha(T_c - T_{c,ref})] \quad (1)$$

Where, $R_{ref}$ the resistance at reference temperature $T_{c,ref}$, usually 20ºC, and $\alpha$ the respective coefficient of the conductor material. The here noted temperature $T_c$ is the average of the strand layers of the conductor across its section [23]. In other words, ambient temperature effects to a conductor and, in turn, to its resistance may be described more accurately with a proper heat exchange model. The 'dynamic case' model of the IEEE 738-2012 standard for DLR [23] is here used. The model is as follows:

$$\frac{dT_c}{dt_T} = \frac{1}{m \cdot C_p}[R(T_c) \cdot I^2 + q_s - q_c(V_w, T_s, T_a,) - q_r(T_s, T_a)] \quad (2)$$

Where $t_T$ the time over which $T_c$ changes, $m \cdot C_p$ total heat capacity of the conductor, $I$ conductor current, $q_s$ heat gain from solar irradiation, $q_c$ convection heat loss and $q_r$ radiated heat loss. It is noted that $q_c$ is a function of wind speed $V_w$, conductor surface temperature $T_s$ and ambient temperature $T_a$, while $q_c$ is a function of $T_s$ and $T_a$. With the heat exchange model in (2), changes in ambient temperature $T_a$ can be expressed in changes to conductor temperature $T_c$ and, thus, in line resistance $R(T_c)$.

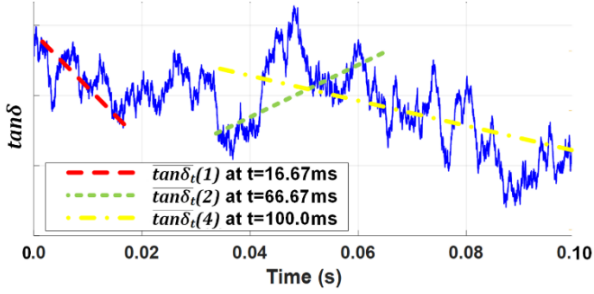

Fig. 2. Depiction of the slope of the moving average of the tangent of the line impedance for various numbers of electrical cycles *k* and times *t*.

To properly account for the changes in ambient temperature induced by a nearby fire seat, we use the model and the parameters for the temperature change $\Delta T_a$ at a location in distance $d$ from a fire seat, in a moss pine forest with several trees and short grass as developed in [28]:

$$\Delta T_a = \frac{q_f(d) \cdot t_F}{A\sqrt{\frac{\pi \cdot \lambda \cdot c \cdot \rho}{4} t_F}} \quad (3)$$

Where $q_f$ thermal radiation from the fire as function of $d$, $t_F$ time of heating, $A$ thermal absorption coefficient per meter of $d$, $\lambda$ thermal conductivity, $c$ specific heat capacity, and $\rho$ air density.

The heat exchange model of the DLR standard in (2) and the one of changes in ambient temperature induced by a fire in (3) are not part of the proposed control method, yet they allow for the analysis of its performance and its tuning, while they also inform a realistic line conductor model for this study. Unlike previous studies that do not account for the realistic effect of ambient temperature to line resistance changes or assume simplistic linear relationships, this study employs the most detailed and practically tested standard on the matter.

*B. Line Disconnection Control due to the Detection of a Fire Through Line Resistance Monitoring*

Continuing from the definition in (1), as resistance $R$ (and, hence, its changes thereof) is typically small in power system lines, the ratio of reactance $X$ over $R$ is a better indicator. It is:

$$Z(T_c) = R(T_c) + jX = |Z(T_c)|\cos\delta(T_c) + j|Z(T_c)|\sin\delta(T_c) \quad (4)$$

Where $Z$ is the line impedance as a function of $T_c$. In other words, the ratio $X/R$ is the tangent of the line impedance angle $tan\delta(T_c)$ when $Z$ is expressed in polar coordinates as in (4). Most typically, overhead conductors in transmission and the medium voltage level of distribution systems are characterized by $X>>R$, which means $tan\delta$ takes large values. Hence, even if $R$ is small, its change will be easier to capture through $tan\delta$. Since the proposed method relies on monitoring changes in $tan\delta$ rather than its value per se, it will be effective even if $X<R$ (i.e. for cases of overhead conductors at the lower voltage levels of distribution system). This is because $tan\delta$ changes at a rate of at least 1/rad (around 0°, i.e. for $X \approx 0$, which is rather unlikely, even if $X<<R$). Another reason why $tan\delta$ is preferred as the control signal, is because there is no meaningful way to directly measure line resistance over any arbitrary segment of one.

Assuming the sending $S$ and receiving $R$ ends of an overhead line (or segment of it) are equipped with PMUs as per Fig. 1, the transferred active $P_R$ and reactive $Q_R$ power to terminal $R$ from terminal $S$ with voltage angle difference $\theta$ between them are described as:

$$\left.\begin{array}{l} P_R = \dfrac{V_R V_S \cos(\delta - \theta) - V_R^2 \cos\delta}{Z} \\ Q_R = \dfrac{V_R V_S \sin(\delta - \theta) - V_R^2 \sin\delta}{Z} \end{array}\right\} \quad (5)$$

Let it be here noted that the line impedance may include shunt capacitance to ground or to neighboring overhead conductors, but by using a pi-model of the line, (5) holds. Testing of the method in Section III assesses this aspect, while it also explores the effects of capacitor bank switching. As long lines are out of the scope of this study, distributed parameter line modeling is not considered. Rearranging and replacing $Z$ within (5) yields:

$$tan\delta = \frac{P_R V_S \sin\theta + Q_R V_S \cos\theta - Q_R V_R}{P_R V_S \cos\theta - Q_R V_S \sin\theta - P_R V_R} \quad (6)$$

If one of the PMUs in the topology of Fig. 1 drives a breaker, the method will control the PMU to open it at the decrease of $tan\delta$; i.e. at the increase of $R$, which will have been caused by increase in ambient temperature, due to a fire near the line. Clearly, $tan\delta$ is not formulated as an estimator as proposed in previous studies, neither does it require system wide measurements. Measurements of the PMUs at the sending and receiving terminals of a line suffice for the control methodology.

To implement the control, $tan\delta$ itself cannot be used as the driving signal, since arbitrary measurement errors can cause unwanted triggering. To this end, the slope of the moving average of $tan\delta$ is preferred, instead. The moving average $\overline{tan\delta}_t(k)$ is calculated over a number of electrical cycles $k$, given a time $t$. At any given time $t$ we define the ratio or slope of the moving average of $tan\delta$ as:

$$\Delta\overline{tan\delta}_t(k) = \frac{\overline{tan\delta}_{t-k\cdot T}}{\overline{tan\delta}_{t-T}} \quad (7)$$

Where $k \in \mathbb{N}_1$, $T$ the period of an electrical cycle and bar denotes the moving average. $\Delta\overline{tan\delta}_t$ will be greater than 1 for an increase in line resistance and less than or equal to 1 otherwise.

In the way we here define $\overline{tan\delta}(k)_t$ and from what was described about the values of $tan\delta$ in (6), $\overline{tan\delta}_t(k)$ as the slope of the moving average $tan\delta$ is expected to be downward to indicate an increase in line resistance. We show in Fig. 2 how $\overline{tan\delta}(k)_t$, the moving average of $tan\delta$, behaves for various numbers of periods $k$ and times $t$. In the depicted example there is an increase in conductor strand layers temperature $T_c$. causing an increase in resistance $R$. Additionally, since the main risk for an overhead line remaining energized during a forest fire is the occurrence of arcing faults [4], $k$ should be selected small enough for fast control action, yet not too small to cause unwanted triggering due to measurement error, it should preferably be $k=$[2-12].

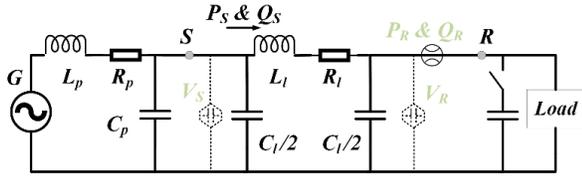

Fig. 3. Tested power system with S-R part of the line monitored with PMUs on both ends. In light fonts are the PMU voltage and current measurements.

TABLE I
TEMPERATURE CHANGE AT DISTANCE $d$ FROM AN OVERHEAD LINE FOR HEATING TIME $t_F$

| $t_F$ (s) | $d$ (m) | $\Delta T_a$ (°K or °C) | $t_F$ (s) | $d$ (m) | $\Delta T_a$ (°K or °C) |
|---|---|---|---|---|---|
| 10 | 50 | 8.23·10⁻⁵ | 10 | 5 | 30.99 |
| 30 |  | 1.42·10⁻⁴ | 30 |  | 53.69 |
| 60 |  | 2.02·10⁻⁴ | 60 |  | 75.93 |
| 10 | 10 | 5.81 | 10 | 1 | 71.61 |
| 30 |  | 10.06 | 30 |  | 124.03 |
| 60 |  | 14.23 | 60 |  | 175.40 |

Multiple studies [29, 30] seek to identify adequate value propositions for the deployment of PMUs. Hence, if an overhead line is disconnected just in time ahead of an approaching fire, it represents a PMU application of notable value. Customers' supply will not be unnecessarily disrupted out of abundance of caution in the wide vicinity of a fire, utilities will not suffer missed opportunity revenues, and unforeseen/unexpected fire seats can be detected via electrical grid monitoring, which is unprecedented. The PMU technology is crucial in this methodology and no other measurement infrastructure could enable it. This is because line disconnections must be implemented promptly (sub-second times) to avoid arcing faults and new fire seats at the fast approach and spread of wildfires. Also, the synchronization of measurements at the line terminals in granularity of ns is a feature that can be widely offered only by PMUs.

## III. TESTING OF THE PROPOSED METHOD AND TUNING

### A. Testing Framework and Results

To test the proposed methodology, we model in MATLAB Simulink the network shown in Fig. 3. It comprises an ideal voltage source $G$ connected to a load via an overhead line. Every conductor (phase) in one part $S$-$R$ of the line is monitored with PMUs per the control described in Section II. Capacitor banks may compensate voltage drops at the load terminals. System ratings are given in the Appendix. The line is assumed to be arbitrarily of any of the ACSR types with $a$=0.004/°C and X/R ratio ranging between 0.13-4.26. Using the heat transfer model in (3) for a fire in distance $d$ from an overhead line, Table I shows some characteristic ambient temperature changes caused by a fire seat.

With all aforementioned models and formulations, we here list the ranges of ambient conditions, system loading, and equipment ratings to test the proposed control method:
i. distances of the fire seat from the overhead line of at most 50m, burning for at least 10s, which, as of Table I, corresponds to $\Delta T_a$ between 0-225.5 °C (including no fire),
ii. ambient temperature $T_a$ ahead of the effect of the fire between 10-40 °C,
iii. wind speed $V_w$ between 0-6.5 m/s,

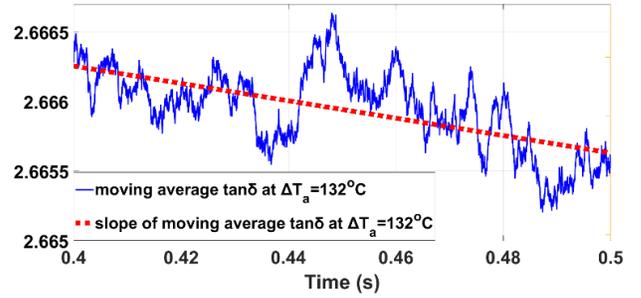

Fig. 4. Downward slope of the moving average of $tan\delta$ of the impedance phasor of a line, which succeeds in detecting a fire near the line causing $\Delta T_a$=132°C.

iv. conductor surface temperature $T_s$ between 10-100 °C,
v. lengths of part $S$-$R$ of the line up to 20km,
vi. line current up to 1600 Amps (considering also step load increase or decrease),
vii. reactive power compensation with switching capacitor banks at the load bus for power factor correction up to 1,
viii. error of magnitude measurement of voltage up to ±0.3% and of current up to ±0.6% owing to the instrument transformers used (see Appendix) and
ix. total PMU vector error up to 1% (see Appendix).

Most of the selected ranges are typical and used in the previously cited works, too, while some others represent worst case scenarios. The PMU total vector error of no more than 1% was taken from [31] for devices of the protection class according to [32]. The effect of current and voltage measurement errors from instrument transformers used is also considered according to IEC and IEEE standards [33, 34], again for the protection class of such components. The PMU voltage angle measurement error at ±0.021° [13] or that from the instrument transformers at ±8'' [33] is negligible considering the lengths of the line and most of the load range tested. The heat from solar irradiation $q_s$ in (2) will be set to zero to represent a worst-case scenario; this is also realistic for cases of forest fires during night times, as also owing to shadowing of the lines by clouds or dense smoke and ash from the fire. The ranges in (i)-(ix) above are split in five intervals each, and every combination among them describes a testing scenario for a total of about 2 million scenarios. For example, one testing scenario is {$\Delta T_a$ = [45.2-90.2) °C, $T_a$ = (34-40] °C, $V_w$ = [2.6-3.9) m/s, $T_s$ = [64-82) °C, $l_{S-R}$ = [12-16) km, $I$ = [0-320) A, $\Delta cos\varphi$ = [0.4-0.6), $V_e$ = [0.0004-0.0006) pu, $I_e$ = [0.002-0.004) pu}. For each scenario more than 16000 tests (statistical validity [35]) of 0.5 s simulation time are conducted on the MATLAB Simulink model of the overhead line supplying power to a load from an ideal voltage source as given in Fig. 3. A modest sampling rate of 5kHz is considered; let it be stressed that every sample could take an arbitrary measurement error within each interval assessed. From each test we collect the time series of $tan\delta$ of the line as calculated by equation (6) and from that we extract $\Delta \overline{tan\delta}_t(k)$ according to equation (7) for $k$=[2-12] and various start times $t$ of calculation within the 0-0.5 s interval of the simulation time.

The expected control action as per the proposed method, is depicted in Fig. 4, where $\overline{tan\delta}_t$ has a downward trend at an increase in ambient temperature $T_a$ causing a change in line resistance captured through $tan\delta$. Although such a trend can be noted in multiple tested cases where $\Delta T_a$>0, it is not consistent, while in some cases the trend is upward. For example, in Fig. 5 we notice such an occurrence, even though there is fire seat causing $\Delta T_a$=29°C. It is evident that the proposed control method will fail to capture some ambient temperature changes caused by a fire, because of any of the following reasons:
- some of the ambient conditions do not allow $\Delta T_a$ to lead to a sufficient change in average temperature of the conductor strand layers $T_c$,
- measurement errors skew the slope $\overline{\Delta tan\delta}_t$,
- the load is very low and/or the monitored S-R part of the conductor is short.

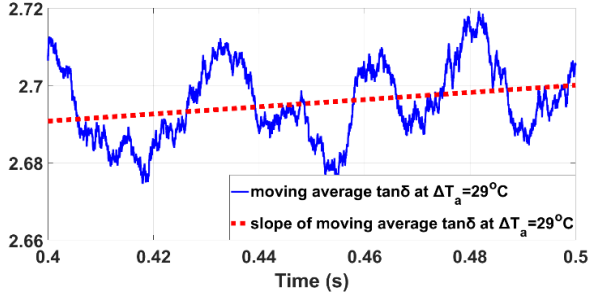

Fig. 5. Upward slope of the moving average of *tanδ* of the impedance phasor of a line, which fails to detect a fire near the line causing $\Delta T_a$=29°C.

These observations may also be confirmed by the contributing factors in the heat exchange model in (2).

*B. Analysis of Results and Tuning of Control Method*

To determine the proper tuning of the control method and under which conditions it best performs its purpose, we analyze the collected test results with binary decision trees (DTs) [36]. DTs are machine learning tools that can extract rules out of data marked as True/False on some statement; these rules describe how the data they represent are noted as True or False. A DT splits the initial dataset in a way that reduces the information entropy of the True/False state in the occurring subsets. The splits are recursively employed in the subsets until a stopping criterion is met. An example of a DT is offered in Fig. 6; the rules are the descriptions of the subsets in the leaves of the DT [37]. For example, leaf 4 reads that 99% of the cases where A7 is False and A1 is less than 8.3, fall under the True state of the classification statement.

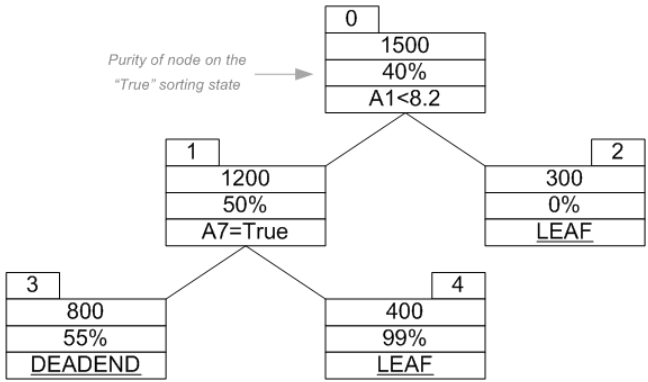

Fig. 6. Example of a binary DT. Reading from the leaves up to the root of the DT defines the rules describing this dataset.

TABLE II
STATISTICS OF TESTING THE TWO TYPES OF CONTROL WITH $\overline{\Delta tan\delta}_t$ UNDER THE SPECIFIED CONDITIONS OF A FIRE SEAT NEAR AN OVERHEAD LINE

| Control type & conditions | $\Delta tan\delta_t$ performance (%) | | | |
|---|---|---|---|---|
| | TP | TN | FP | FN |
| Control 1 with $\Delta T_c$>2.87°C | 99.32 | 0.29 | 0.29 | 0.10 |
| Control 2 with $\Delta T_c$>2.87°C and $V_{err}$<0.003% | 89.13 | 0.00 | 0.00 | 10.87 |

For all scenarios of the ranges (i)-(ix) we train a DT after labeling all tests where $\Delta T_a$>0 as True and the rest as False. We seek to extract rules that describe a control action that successfully detects a fire in more than 90% of the cases of that testset. The Shannon entropy information gain ratio splitting criterion is used [37].

We find that two types of control with the proposed method may detect a fire near an overhead conductor line. These are:

Control 1: $\Delta\overline{tan\delta}_t(6) > 1$ AND
  ( $\Delta\overline{tan\delta}_t(4) > 1$ OR $\Delta\overline{tan\delta}_t(3) > 1$ )

Control 2: $\Delta\overline{tan\delta}_t(6) > 1$.

Control 1 detects a fire if the line resistance has been increasing by average (moving average of *tanδ*) in the past 6 consecutive periods and in either the last 4 or last 3 consecutive periods. Control 2 detects a fire if the line resistance has been increasing by average (moving average of *tanδ*) in the past 6 consecutive periods. The two control tunings perform their purpose under two different sets of conditions (specific intervals of the ranges of weather, loading and measurement errors). *Control 1* may be best used when $\Delta T_c$>2.87°C, while *Control 2* should be preferred when $\Delta T_c$>2.87°C and the PMU voltage measurement error is less than 0.00003 pu. As it can be noted, these controls detect a nearby fire from changes to line characteristics particularly fast, in time as little as a few ms. In this test set-up, just 6 electrical cycles are enough for the purpose.

It is reminded that $\Delta T_c$ is the change of average temperature of the conductor strand layers. If equipment relevant to DLR is not used [25], $\Delta T_c$ cannot meaningfully drive the application of the proposed control methodology. To this end, DTs are used anew to relate $\Delta T_c$ to ambient and loading conditions. To do so we group all tests sets from all scenarios of ranges (i)-(ix) and label as True those tests with $\Delta T_c$>2.87°C and False otherwise. We find that $\Delta T_c$>2.87°C occurs under the following conditions:
a. 86% of the cases with $\Delta T_a$>76°C,
b. 100% of the cases with $V_w$<1.35m/s and $|P_R+jQ_R|$>90% of the line static rating, and
c. 94% of the cases with $T_s$<57°C, $|P_R+jQ_R|$>50% of the line static rating and $\Delta T_a$>46°C.

For (c), it is noted that the conductor surface temperature can be estimated from weather conditions per a heat exchange model [23, 24]. Following this analysis, we may now state that:
- *Control 1* will detect an approaching fire if the ambient temperature change is greater than 76°C (i.e. fire burning for at least 60 s at a distance of at most 5 m from the conductor) *or* if the wind speed is less than 1.35m/s and the line is loaded above 90% of its nominal capacity *or* if conductor surface temperature is below 57°C, the ambient temperature change is above 46°C (i.e. fire burning for at least 10 s at a distance of at most 10 m from the conductor) and the line is loaded above half its nominal capacity,
- *Control 2* will detect an approaching fire under the same conditions as *Control 1*, but for PMU voltage measurement error not exceeding 0.00003 pu of rated/monitored grid voltage.

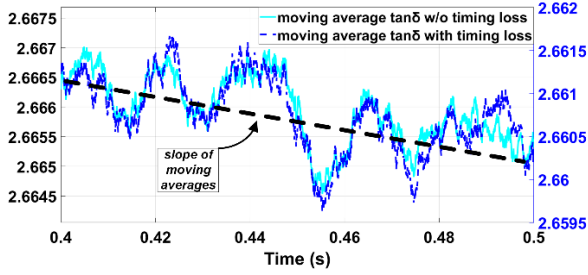 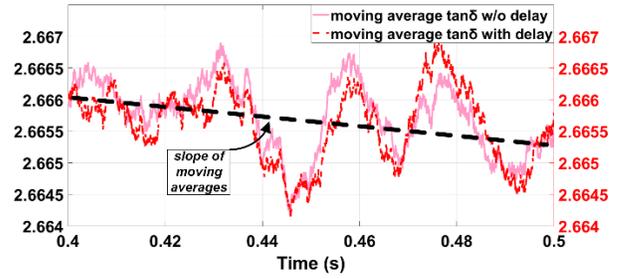

Fig. 7. Downward slope of moving averages of *tanδ* of the impedance phasor of a line, without and with timing loss of one of the PMUs of the method.

Fig. 8. Downward slope of moving averages of *tanδ* of the impedance phasor of a line, without and with timing delay of one of the PMUs of the method.

All other conditions (not mentioned) need not be met, may vary freely within the intervals mentioned in (i)-(ix) in Section III.A and will not affect the control performance. For example, the PMU voltage and current measurement errors for *Control 1* may be as high as 0.001 pu and 0.01 pu, respectively.

We assessed the performance of *Control 1* and *Control 2* under the respective conditions they perform best, to determine whether the controls would correctly detect a fire seat near an overhead line or mistakenly cause undesirable disconnection. The performance is described in terms of true positive (TP), true negative (TN), false positive (FP) and false negative (FN) action of the controls. The results are summarized in Table II.

Let it be here stressed that DTs were preferred as typical off-the-shelf solutions [36] and are not the only classification tool that may be used for the purpose. Regression, neural networks, support vector machines and others [38] could equally serve the requirements to determine the control formulation and the identification of the conditions of their best performance thereof.

### C. Effects of Timing Loss

Loss of synchronization of PMU measurements [39] is a source of concern, as it introduces non-trivial/standardized error. For this reason, we explore two cases of timing loss: a persistent timing loss and a delay. A persistent timing loss means that either of the PMUs in the tested set-up of Fig. 3 reports a constant measurement following a permanent disruption to its synchronizing clocking. A delay means that either of the PMUs in the tested set-up reports a measurement that is out-of-synchronization with the other PMU by a given amount of time.

We test the two cases of PMU timing loss for *Control 1* under random instances of condition (b) as reported in the previous subsection. Condition (b) is the worst-case scenario for testing timing loss, because we expect 100% successful detection of a forest fire in the vicinity of the monitored conductor. The TP outcome would be that of Fig. 4, while the undesired FN of Fig. 5 would mean that the timing loss affects the performance of the method. We stress that we assess the effect of timing loss that occurs in the middle of a fire approaching the conductor and in the effort of our method to detect it. The reason for that is that there have already been techniques [40, 41] proposed to mitigate timing loss and compensate synchronization for time-frames longer than the few electrical cycles of our method. At no effect to generality and by design, we assume the timing loss cases occurring at the sending terminal $S$ of the testing set-up. For the delay, 0.1 s of delay is examined.

As it may be noted in both Fig. 7 and Fig. 8, the slope of the moving average of *tanδ* is unaffected by either a permanent timing loss or a delay, even though some differences in absolute *tanδ* value are seen. This should be expected as of (6) because:
- without steep load changes (which are also slower mechanical phenomena), angle $θ$ will be unaffected, and
- as an approaching fire increases conductor resistance, $V_S$ and $V_R$ will both drop, hence, one of them "appearing" constant would not affect the slope of the moving average of *tanδ*.

As of the above we conclude that the proposed methodology can detect an approaching fire front to an overhead conductor, even if timing losses occur in the midst of the event.

## IV. DISCUSSION AND IMPLEMENTATION

### A. Comparing Testing and Field Conditions

The potential of the proposed control methodology according to the results in Section III must be considered within the framework of testing under worst-case scenario conditions. Most notably, these conditions assume that, firstly, the conductor surface temperature $T_s$ may be as high as 100ºC, which is unlikely, given that this is the operating temperature limit for such lines [42, 43]. Secondly, the tests completely discounted the effect of solar irradiation that would further increase the temperature differential over time from (2) and, thus, ease its capture from the control signal $\overline{Δtanδ}_t$. The reason why these conditions were chosen, was because of the lack of proper real-word testing (see Conclusion for future directions of this work).

That been said, it is found that for fire seats burning at a distance of about 5-10 m from an overhead conductor, the proposed method may correctly detect the fire in at least 85% of all cases of average line loading or of low wind speed. This percentage represents the worst-case scenario for TP outcomes of *Control 1* (see Table II) under the 86% of conditions when the ambient temperature is greater than 76ºC (i.e. fire burning for at least 60 s at a distance of at most 5 m from the conductor). Under the same conditions, the method will fail to detect the fire in no more than 10% of the cases, while an unwanted disconnection will happen in less than 0.5% of all occurrences. Extending the testing with more favorable than worst-cases scenario assumptions will yield additional conditions for effective use of the method.

In a large forest fire with multiple fire seats, such a method can automate line disconnections, provided PMUs are installed at these lines and in some proximity among them. A few CA utilities have already deployed PMUs along their overhead conductors in distances ranging as short as 1-3 miles (info from Acknowledgement). The testing in Section III included scenarios of line lengths as short as these. Another interesting aspect here is that many substations are already equipped with PMUs, hence, equipping the lines directly connected to them with PMUs, represents a strategy of immediate applicability.

*B. Parametrizing Control*

We have earlier proposed to tune the control of the methodology with DTs (or other kind of classification or analytics tool). However, actual data of fires near lines might not be available for the purpose. We here note that for any line sought to be monitored and protected from the approach of a fire with this methodology, there will be data about the weather conditions in the area, the composition of the fuel for a potential fire nearby (for the parameters of the temperature change $\Delta T_a$ in (3)) as also the loading levels and characteristics of that line. Many utilities already assess conductor temperatures using DLR measurement data [25, 44], that can be sourced for this method.

With this information, Monte Carlo simulations may be conducted to generate synthetic, yet realistic, data, which can, in turn, be used to train the classification or analytics tool and determine the PMU controls for the purpose of this method. Monte Carlo simulations to generate realistic data is not an uncommon practice for utilities, considering, for example, the typical problem of hydro-thermal scheduling [45, 46].

*C. Equipment and Infrastructure Concerns*

In many cases we have to also consider the switching effects from series compensation within the *S-R* part of the monitored line, which would affect $\Delta \overline{tan\delta}_t$. In such a case and assuming a fire burns in the vicinity of the conductor, the moving average formulation would still present a downward trend even if it would transiently suffer a step change. At the activation or de-activation of some series compensation, $\Delta \overline{tan\delta}_t(6)$ (necessary for both *Control 1* and *Control 2*) would restart with $k=1$, until $\Delta \overline{tan\delta}_t(6) > 1$ occurs again and indicates the nearby fire. In other words, any series compensation switching will delay the disconnection of the line by a maximum of 6 electrical cycles.

Latencies from measurement reporting and communication may also affect the effectiveness of the methodology. Since PMUs must abide by typical industry standards [32, 47], delays in the calculation of $\Delta \overline{tan\delta}_t$ will be within well-defined limits. For example, for a PMU of the protection class reporting measurements at a rate of 10Hz (which coincides with the time interval of $\Delta \overline{tan\delta}_t(6)$), standard [32] allows for a reporting latency of no more than 200 ms, while the communication latency (to collect the measurements of the PMU at the other end of the monitored line) would typically not exceed 35 ms [47]. In the worst-case scenario, these delays would add up to 235 ms and the total time to detect the nearby fire through $\Delta \overline{tan\delta}_t(6)$ plus the noted delay, would not exceed 0.5 s.

## V. Conclusion

This work presents a control method for the disconnection of an overhead line just in time as a fire-front approaches or breaks outs. The method is uniquely enabled thanks to the use of synchronized phasor measurements at the ends of a line or a segment of it equipped with PMUs. The method relies on the increase of line resistance as the ambient temperature increases. This line resistance increase is captured via the tangent of the line impedance, as it is easier and more accurate to measure through synchrophasors.

A detailed model accounting for the thermal dynamics of an overhead conductor at the vicinity of a fire seat were used to properly account for realistic scenarios, albeit they were preferred to be set up as worst-cases. The exhaustive testing conducted considered broad ranges of weather conditions and PMU measurement errors. The results, after analyzed with binary decision trees, were found to lead to two control formulations that correctly detect a fire near an overhead line with high degree of certainty under some conditions. It has been shown that the method will accurately capture temperature changes due to a fire at least 5 m from an overhead line and for burning times of at least 10 s, while it is set up to disconnect the line in a time of six electrical cycles to protect it against arcing faults.

Further research plans of the Authors include a hardware-in-the-loop set-up with PMU devices and collaboration with the fire laboratories of the US Forest Service, for a more pragmatic assessment of the performance of the technique. Additional testing of the methodology could consider more favorable testing conditions that would only yield additional control formulations for more conditions of a fire seat near a line.

Combining the method with control techniques that detect down conductors could also be a valuable research direction.

Since dynamic thermal rating models have also been standardized for transformers, the framework of this method could also be extended to the protection of substations in areas susceptible to forest fires. In this sense, the substations will be de-energized just in time and not preemptively. The value proposition of this research extension could be considered much greater than that of the proposed control for lines, as a substation represents more load demand than what any single overhead line does.

## VI. Appendix

<u>Overhead conductor characteristics</u>:
Type: ASCR,
Rated Operating Voltage = 138kV,
Rated Current = 2.1kA,
X/R ratio = [0.13-4.26]

<u>Instrument transformer characteristics</u> [33, 34]:
Voltage magnitude error = ±0.3%,
Current magnitude error at $I > 20\% \cdot I_n$ = ±0.3%
Current magnitude error at $I \leq 20\% \cdot I_n$ = ±0.6%

<u>PMU accuracy characteristics</u> [32]:
Total Vector Error < 1%.

## VII. ACKNOWLEDGEMENT

The Authors wish to acknowledge the contribution of Mr Daniel Dietmeyer co-chair of Distribution Task Team at the North American Synchro-Phasor Initiative (NASPI) in identifying deployments of PMUs at overhead conductors in CA.

The Authors wish to acknowledge the contribution of Dr. Farnoosh Rahmatian in identifying the instrument transformer classes (and the respective standards defining them) that would best serve the purpose of the method.